\documentclass[sigconf]{acmart}

\usepackage{caption}
\usepackage{soul}
\usepackage{algorithm}
\usepackage{algpseudocode}
\usepackage{subcaption} 
\usepackage{tabularx}
\usepackage{booktabs}
\usepackage{listings}
\usepackage{framed}
\usepackage{fontawesome}
\usepackage[english]{babel}
\usepackage[autostyle, english = american]{csquotes}

\MakeOuterQuote{"}

\definecolor{myDarkGreen}{HTML}{b9e192}
\definecolor{myLightGreen}{HTML}{cfebb6}
\definecolor{myDarkBlue}{HTML}{8383F6}
\definecolor{myLightBlue}{HTML}{b3c7f7}
\colorlet{shadecolor}{myLightBlue!15!white}

\newcommand{\rqsummary}[2]{%
\par\smallskip
  \noindent
  {%
    \setlength{\fboxrule}{0.5pt}%
    \setlength{\fboxsep}{6pt}%
    \fcolorbox{myLightBlue!50!black}{myLightBlue!15!white}{%
      \parbox{\dimexpr\linewidth-2\fboxsep-2\fboxrule\relax}{%
        \textbf{Summary of #1.} \textit{#2}%
      }%
    }%
  }%
}

\AtBeginDocument{%
  }

\copyrightyear{2026}
\acmYear{2026}
\setcopyright{cc}
\setcctype{by}
\acmConference[IUI '26]{31st International Conference on Intelligent User Interfaces}{March 23--26, 2026}{Paphos, Cyprus}
\acmBooktitle{31st International Conference on Intelligent User Interfaces (IUI '26), March 23--26, 2026, Paphos, Cyprus}
\acmPrice{}
\acmDOI{10.1145/3742413.3789148}
\acmISBN{979-8-4007-1984-4/2026/03}

\begin{document}

\title{Developer Interaction Patterns with Proactive AI: A Five-Day Field Study}

\author{Nadine Kuo}
\authornote{Both authors contributed equally to this research.}
\orcid{0000-0002-6779-8454}
\affiliation{%
  \institution{JetBrains}
  \city{Amsterdam}
  \country{Netherlands}
}
\email{nadine.kuo@jetbrains.com}

\author{Agnia Sergeyuk}
\authornotemark[1]
\orcid{0009-0001-1495-9824}
\affiliation{%
  \institution{JetBrains Research}
  \city{Belgrade}
  \country{Serbia}
}
\email{agnia.sergeyuk@jetbrains.com}

\author{Valerie Chen}
\orcid{0009-0007-2783-0265}
\affiliation{%
  \institution{Carnegie Mellon University}
  \city{Pittsburgh, PA}
  \country{United States}
}
\email{vchen2@andrew.cmu.edu}

\author{Maliheh Izadi}
\orcid{1234-5678-9012}
\affiliation{%
  \institution{Delft University of Technology}
  \city{Delft}
  \country{Netherlands}
}
\email{m.izadi@tudelft.nl}


\begin{abstract}
Current in-IDE AI coding tools typically rely on time-consuming manual prompting and context management, whereas proactive alternatives that anticipate developer needs without explicit invocation remain underexplored. Understanding when humans are receptive to such proactive AI assistance during their daily work remains an open question in human-AI interaction research. We address this gap through a field study of proactive AI assistance in professional developer workflows.
We present a five-day in-the-wild study with 15 developers who interacted with a \textit{proactive} feature of an AI assistant integrated into a production-grade IDE that offers code quality suggestions based on \textit{in-IDE} developer activity. We examined 229 AI interventions across 5,732 interaction points to understand how proactive suggestions are received across workflow stages, how developers experience them, and their perceived impact.
Our findings reveal systematic patterns in human receptivity to proactive suggestions: interventions at workflow boundaries (e.g., post-commit) achieved 52\% engagement rates, while mid-task interventions (e.g., on declined edit) were dismissed 62\% of the time. Notably, well-timed proactive suggestions required significantly less interpretation time than reactive suggestions (45.4s versus 101.4s, $W=109.00$, $r=0.533$, $p=0.0016$), indicating enhanced cognitive alignment.
This study provides actionable implications for designing proactive coding assistants, including how to time interventions, align them with developer context, and strike a balance between AI agency and user control in production IDEs.
\end{abstract}



\begin{teaserfigure}
  \includegraphics[width=\textwidth]{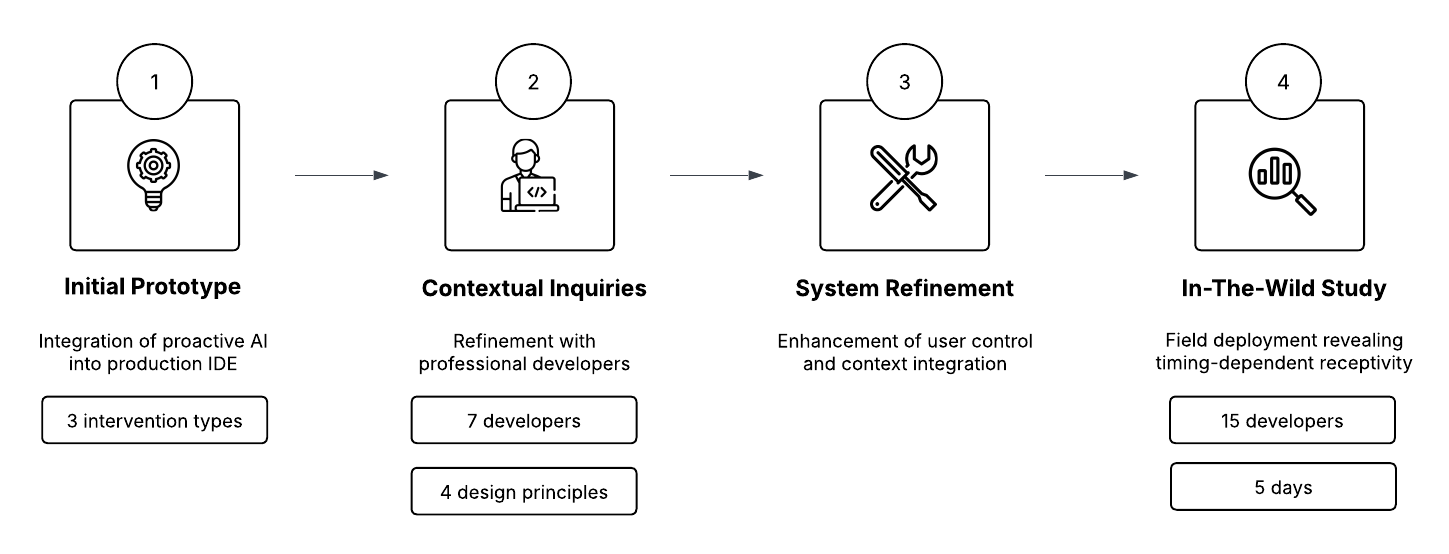}
  \caption{\textbf{Study Overview.} We developed and iteratively refined \emph{ProAIDE}, a proactive feature for code quality of an in-IDE AI assistant (Phases 1-3), then conducted a 5-day field study with 15 professional developers (Phase 4).  Analysis of 229 interventions across 5,732 interaction points revealed that timing critically impacts receptivity: post-commit suggestions were more readily accepted than mid-task interventions.}
  \Description{A horizontal flowchart showing four phases of the study. Phase 1 (Initial Prototype): Integration of proactive AI into production IDE, with 3 intervention types. Phase 2 (Contextual Inquiries): Refinement with professional developers, 7 developers, 4 design principles. Phase 3 (System Refinement): Enhancement of user control and context integration. Phase 4 (In-The-Wild Study): Field deployment revealing timing-dependent receptivity, 15 developers, 5 days. Arrows connect the phases from left to right, with simple icons above each phase representing the activity type.}
  \label{fig:teaser}
\end{teaserfigure}


\maketitle

\section{Introduction}

Programming tools powered by large language models (LLMs) have changed how developers interact with code in IDEs~\cite{dell2025cybernetic, tabarsi2025llms, weisz2022perfection, jiang2022syntax, izadi2024language}. 
These tools evolved from simple code completions~\cite{microsoft2023intellisense,tabnine} to collaborative assistants that support various software development tasks~\cite{github2021copilot,jetbrains2025junie,ionescu2025multi}. 
Despite these advancements, existing LLM-based tools often face practical limitations when adopted in real-world developer workflows~\cite{sergeyuk2025humanaiexperienceintegrateddevelopment, barke2023grounded, mozannar2024realhumanevalevaluatinglargelanguage}. 
Most tools require explicit prompts from developers, creating cognitive overhead through the need to craft precise queries, supply context, interpret responses, and integrate suggestions~\cite{mozannar2024readinglinesmodelinguser, weisz2022perfection, jiang2022syntax}. 
These costs diminish productivity gains, especially in complex or urgent tasks, where frequent context switching between coding, prompting, and interpreting outputs interrupts developers’ flow and increases mental fatigue.

To reduce such cognitive overhead, both researchers and practitioners have begun exploring proactive AI programming assistance, capable of anticipating developer needs and offering suggestions without explicit requests~\cite{microsoft2018intellicode,github2021copilot,cursor2023,windsurf2023,sergeyuk2024designspaceinidehumanai, chen2025needhelpdesigningproactive,pu2025assistance}.  
Yet, a key challenge persists: when prompts are poorly timed or perceived as intrusive, they risk undermining the user experience. Existing results highlight a critical open problem: how to accurately identify suitable moments for intervention that support, rather than disrupt, the user's workflow. Moreover, existing evaluations of proactive assistants have been confined to \textit{controlled}, \textit{web-based} environments with \textit{predefined} tasks and \textit{monolingual} settings~\cite{chen2025needhelpdesigningproactive,pu2025assistance}. These limitations restrict our understanding of how proactive assistants perform under real-world conditions, where timing, contextual relevance, and code maintainability are critical concerns.

In this work, we investigate how proactive capabilities can be integrated into in-IDE assistants, advancing toward practical, context-aware developer support. 
We focus on providing proactive assistance to improve \emph{code quality}, a particularly relevant developer concern that often goes unnoticed during development~\cite{yamashita2013developers} despite its direct impact on code review efficiency, production defect rates, and deployment velocity~\cite{bacchelli2013expectations,nagappan2006mining,tornhill2022code}. 
While developers have shown strong interest in AI tools that help with code quality, including bug detection, refactoring, and adherence to best practices~\cite{chen2024impactaipaircodequality, tan2024faraipoweredprogrammingassistants, chen2025needhelpdesigningproactive}, many existing tools addressing code quality are offline or reactive~\cite{wadhwa2024core, bouzenia2025repairagent, jaoua2025combininglargelanguagemodels}.
Reactive tools provide limited usefulness in fast-paced development settings where timing and context are critical~\cite{xu2025mantraenhancingautomatedmethodlevel, wu2024ismell}. 
We aim to develop proactive assistance that is timely and non-obtrusive to help developers resolve problems before they might escalate. 

To this end, we design, implement, and evaluate through a field study \emph{ProAIDE}, a proactive feature of an AI assistant, that offers suggestions for code quality improvements. ProAIDE is integrated into Fleet\footnote{Fleet: \url{https://www.jetbrains.com/fleet/}}, a production-grade, polyglot IDE with built-in AI capabilities used by professional developers, and owned by JetBrains\footnote{JetBrains: \url{https://www.jetbrains.com/}}, a major software vendor specializing in the creation of intelligent development tools.
The development of ProAIDE followed an iterative, human-centered design process comprising four phases (Figure~\ref{fig:teaser}): in Phase 1, we built an initial prototype with three trigger types for proactive interventions based on established guidelines~\cite{horvitz1999principles,amershi2019guidelines,chen2025needhelpdesigningproactive,pu2025assistance}; in Phase 2, we conducted contextual inquiries with seven professional developers to refine the system's timing heuristics and interface design, extracting four design principles; in Phase 3, we refined the system based on these insights, resulting in the final ProAIDE implementation; and, in Phase 4, we deployed ProAIDE in a five-day in-the-wild study with 15 professional developers working on authentic development tasks spanning 12 programming languages. 
This integration enabled an in situ evaluation of proactive support in realistic workflows. 
Our analysis takes a mixed-methods approach, combining telemetry logs with structured daily and post-study surveys.
This setup allowed us to analyze not only how developers interact with in-IDE proactive AI suggestions, but also how developers experience these interactions and the impact of proactive assistance on their work.

Our results offer a deeper understanding of how, when, and why proactive suggestions can support adoption and usefulness in real-world developer workflows. Across $5,732$ interaction points, we found that suggestions triggered at natural workflow boundaries, particularly after commits, were significantly more likely to be accepted. 
On a System Usability Scale (SUS)~\cite{lewis2018system}, developers rated ProAIDE as easy to use (SUS = 72.8 out of 100, 95\% CI [64.1, 81.5]) and appreciated its integration into the IDE, but varied in their preferences for timing, control, and granularity of suggestions. 
While alignment with task intent shaped trust and receptiveness, utility declined when the AI failed to capture lower-level contextual understanding of the developer's source code, including technical design choices or domain-specific patterns.
Compared to reactive use of AI, proactive suggestions were interpreted significantly faster ($W=109.00$, $r=0.533$, $p=.002$), revealing the potential for enhanced workflow efficiency.
Based on our empirically grounded findings, we present design implications for integrating proactive AI assistance with real-world IDEs to guide future systems toward more developer-aligned support. 

Our contributions are as follows.
\begin{itemize}
    \item \textbf{Empirical characterization of developer interaction patterns} with proactive AI suggestions across workflow contexts, revealing timing-dependent receptivity patterns in authentic professional environments.

    \item \textbf{Quantitative evidence of proactive assistance efficiency}, showing significantly reduced interpretation time compared to reactive AI use through deployment in production IDE.

    \item \textbf{Empirically grounded implications for building} proactive assistants: when to surface suggestions, what makes them effective, how trust and context influence adoption.

    \item We share our study material via a \textbf{replication package}\footnote{We share recruitment texts, survey instruments, and contextual inquiry protocol. Note that the collected user data cannot be made public due to user privacy considerations.}~\cite{supplementary2025}.
\end{itemize}
\section{Background}

\subsection{Developer-AI Interaction and Proactive Assistance}
Understanding how developers interact with \textit{proactive} AI assistants in the real-world IDE is lacking in current human-AI interaction research, compared to developer interactions with \textit{reactive} systems.
For example, \citet{ross2023programmer} demonstrated the value of extended conversational interactions with AI systems for complex tasks, showing that multi-turn dialogues can enhance productivity compared to single-invocation approaches. Relatedly, \citet{barke2023grounded} showed through a grounded theory that user interactions with AI can be classified into two modes akin to the two systems of thought in dual-process theories of cognition: acceleration and exploration. 
For interactions with reactive assistants, \citet{mozannar2024reading} suggested categorization of 12 programming states, including the most prominent time-wise verifying suggestions, prompt crafting, and editing code. 
In the recent work of \citet{sarkar2025vibe}, it was found that vibe coding with AI follows cycles: set a goal, prompt the AI, test results, refine, and sometimes edit manually. 
As these works primarily investigate \textit{reactive} scenarios where humans explicitly seek AI assistance, they place the cognitive burden of recognizing assistance opportunities entirely on the developer. This interaction model introduces cognitive load and often limits real-time utility~\cite{chopra2023conversationalchallengesaipowereddata, mozannar2024realhumanevalevaluatinglargelanguage, nam2024usingllmhelpcode}. 

Proactive tools take a different approach. These systems anticipate developer needs and offer assistance without the need for explicit prompting. However, a fundamental challenge persists: determining when AI systems should \textit{proactively} intervene without disrupting developers' workflow and cognitive state~\cite{horvitz1999principles,amershi2019guidelines}. The timing challenge is particularly critical in real-world professional developers' workflow because poorly timed interventions can disrupt focused work and reduce trust, while well-timed assistance can enhance workflow efficiency without cognitive overhead~\cite{bailey2000measuring, iqbal2005towards}. 

While research on more proactive assistants has been studied in forms ranging from physical robots~\cite{baraglia2016initiative,peng2019design} to virtual chat assistants~\cite{liao2016can,liu2024compeer}, the actual capabilities of prior deployed proactive assistants do not always meet user expectations~\cite{meurisch2020exploring}. 
Moreover, the effectiveness of proactive AI in professional software development contexts remains underexplored. 
Only some research investigated how to improve the timing and scope of proactive code completions~\cite{de_Moor_2024, Mozannar2023WhenTS} and several works explored proactive chat-based agents for professional coding~\cite{chen2025needhelpdesigningproactive, zhao2025codinggenieproactivellmpoweredprogramming}. 
As many of the previous systems relied on simple models or preset messaging, with the advent of modern LLMs and their growing usage in agents for software engineering~\cite{yang2024swe}, there is significant potential to design more capable proactive coding assistants. 

Furthermore, a critical limitation of existing proactive AI research in software development is the reliance on controlled evaluation environments.
Most of these systems were evaluated in toy editors, controlled settings, or Python-only environments~\cite{chen2025needhelpdesigningproactive, pu2025assistance}. 
These design considerations limit what we know about how proactive assistance behaves in full-featured IDEs with realistic workflows, diverse programming languages, and authentic time pressures. 
Building on prior work, we integrate proactive support into a production-ready in-IDE AI assistant and conduct a five-day in-the-wild deployment with professional developers. Through this deployment, we capture developer interactions with proactive suggestions triggered at three AI-driven workflow-grounded points~\cite{bradley2022sources, chattopadhyay2019latent}: ambiguous prompt detection, declined AI edit follow-up, and post-commit review.

\subsection{Code Quality as a Domain for Proactive Assistance}
Developers increasingly expect AI tools to support not just their productivity (e.g., how quickly they can write code), but also make meaningful edits to existing code to improve the code quality~\cite{bird2023taking, sergeyuk2025using}. 
Poor code quality has direct business consequences: it increases time spent in code review~\cite{bacchelli2013expectations}, leads to higher defect rates in production~\cite{nagappan2006mining}, and reduces deployment frequency~\cite{tornhill2022code}. Despite these impacts, code quality issues frequently escape detection during initial development~\cite{yamashita2013developers}, making them ideal candidates for proactive AI intervention.
While recent work shows that LLMs can help with refactoring~\cite{Pomian_Bogolov_emassist_2024}, bug fixing~\cite{wadhwa2024core, bouzenia2025repairagent}, review~\cite{cihan2025evaluatingcodereviewllms}, optimization~\cite{gao2024searchbasedllmscodeoptimization}, and smell detection~\cite{wu2024ismell}, the aforementioned tools are largely reactive.
Developers must spot a problem, switch tasks, and phrase a suitable prompt, which are multiple barriers that can limit adoption.

Proactive support can reduce the friction of reactive tools by offering timely, relevant improvements without requiring initiation. 
Moreover, code quality suggestions often involve changes that do not affect program behavior, such as style improvements or small refactorings. 
These suggestions can still be helpful even when they are not directly related to the developer's current task. 
Developers also appreciate support that enhances clarity and maintainability, as long as it is not disruptive~\cite{chen2025needhelpdesigningproactive, pu2025assistance}. 
Our work builds on these observations by exploring how proactive AI can surface code quality improvements in real workflows. 
Note that we do not focus on building new quality assessment tools and instead study how such improvements can be integrated into natural developer workflows.
\section{Proactive System Design and Iterative Refinement}\label{sec:design-principles}

\begin{figure*}[tbp]
    \centering
    \includegraphics[width=\textwidth]{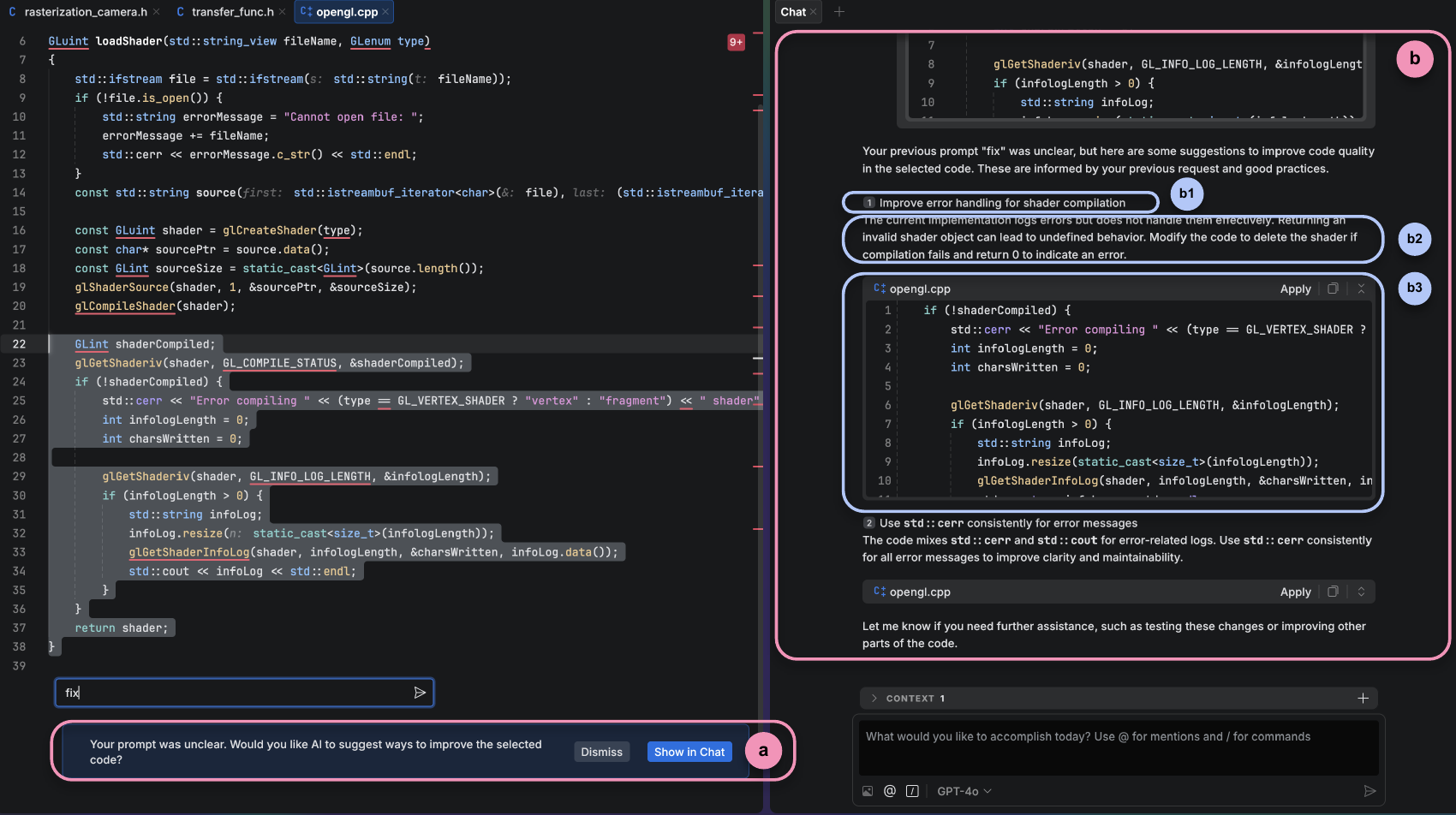}
    \caption{\textbf{User interface for proactive intervention at an \textit{ambiguous prompt}.} (a) Lightweight in-editor cue indicating AI activity. 
    (b) Opened chat panel with: (b1) suggestion title, (b2) rationale, and (b3) code snippet.
    The UI for \textit{declined AI edit} followed the same structure. For \textit{post-commit}, the interaction interface remained, but the visual cue (a) appeared as a popup in the upper-right corner after a successful commit.}
    \Description{Screenshot of the IDE interface showing proactive intervention for an ambiguous prompt. The left side shows a code editor with C++ code. At the bottom, a lightweight banner reads "Your prompt was unclear. Would you like AI to suggest ways to improve the selected code?" with Dismiss and Show in Chat buttons. The right side shows an opened chat panel containing: a suggestion title about improving error handling, a rationale explaining the issue, and two code snippets with Apply buttons showing the suggested improvements for error handling.}
    \label{fig:ui}
\end{figure*}

To investigate how professional developers interact with proactive AI assistance across workflow contexts in situ, we created a prototype of a feature that provided proactive suggestions for code quality improvements inside a production in-IDE AI assistant. We developed ProAIDE through an iterative, human-centered design process spanning three phases (Figure~\ref{fig:teaser}). First, we created an initial prototype with three workflow-grounded intervention triggers based on prior work (Section~\ref{sec:initial-prototype}). Second, we conducted contextual inquiries with seven professional developers to refine the system and extract four design principles (Section~\ref{sec:contextual}). Third, we refined the prototype based on these insights, resulting in the final ProAIDE implementation (Section~\ref{subsec:system}).

\subsection{Initial Prototype}\label{sec:initial-prototype}

For the initial user interface, the in-IDE's AI chat served as the central communication channel (see Figure~\ref{fig:ui}), specifically designed to minimize context switching while preserving developer focus and workflow continuity based on guidelines from prior work~\cite{vaithilingam2023towards,weisz2023toward,chen2025needhelpdesigningproactive, pu2025assistance}.
When the system detected intervention opportunities, which are described below, it displayed a minimal inline banner providing options to dismiss or engage with the suggestion. 
Upon engagement, the chat panel presented suggestions with associated code patches. For automatically integrating suggestions, diffs were directly displayed in the editor for review. Developers could subsequently accept or reject changes. The initial interface remained virtually unchanged in the final system, with the exception of detailed specifications and interaction flows described in Section~\ref{subsec:system}.

\begin{figure*}[tbp]
    \centering
    \includegraphics[width=.7\textwidth]{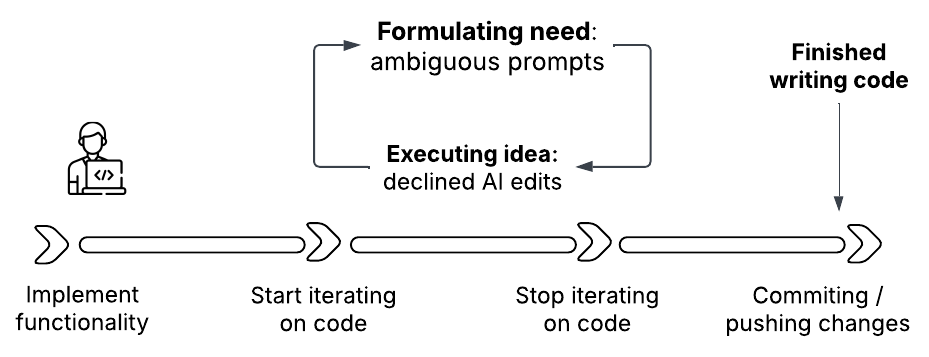}
    \caption{Main stages of the code iteration workflow and corresponding points where ProAIDE triggers proactive suggestions.} 
    \Description{A workflow diagram showing the code iteration process with three intervention points. The flow moves from left to right: Implement functionality leads to Start iterating on code, which branches to two intervention points (Formulating need: ambiguous prompts, and Executing idea: declined AI edits), then continues to Stop iterating on code, and finally Committing/pushing changes where the third intervention point occurs (Finished writing code). The diagram shows how ProAIDE triggers proactive suggestions at these three distinct workflow stages.}
    \label{fig:proactive_timing}
\end{figure*}

Unlike prior work, our in-IDE assistant is present throughout the entire developer workflow. 
In particular, the system triggers proactive suggestions across three conceptual stages of the code improvement process, depicted in Figure~\ref{fig:proactive_timing}. These trigger points were informed by prior studies on developer workflow patterns~\cite{bradley2022sources, chattopadhyay2019latent} and were refined through contextual inquiries: 

\begin{itemize}\label{sec:stages}
    \item \textbf{Formulating Need}: When developers submitted a prompt through the IDE's built-in inline AI action ``Edit Code'', the system evaluated this request by making LLM-as-judge assessments against predefined criteria to identify \textit{ambiguous prompts}, such as lack of context (e.g., requesting a fix without providing error details) or inconsistent references (e.g., using undefined variables). 
    Upon detecting such ambiguity, the system intervened to offer additional guidance and code quality targeted suggestions to enhance their code further. 
    This heuristic is motivated by prior work showing that accurately formulating specifications is non-trivial, imposes substantial cognitive overhead, and has a significant impact on both the efficiency and quality of generated code~\cite{nam2024usingllmhelpcode, yu2025spec2rtlagentautomatedhardwarecode}.
    
    \item \textbf{Executing Idea}: Developers \textit{declining AI code edits} was treated as a heuristic signal that the proposed solution may not have fully aligned with their intent. To explore whether follow-up assistance could be valuable, ProAIDE proactively offered suggestions in such cases with focus on code quality. These could include the reasoning behind the original solution or alternative implementation approaches, aiming to reduce the need for developers to reformulate requests after an unsuccessful attempt.

    \item \textbf{Getting Feedback on Written Code}: the system also proactively provided suggestions after developers \emph{committed their changes}, representing a natural checkpoint~\cite{iqbal2005,pu2025assistance}. Committing code generally indicates that a developer has finished iterating on a logical unit of work, potentially creating a receptive moment for review. Hence, upon successful commits, developers were proactively presented with an option to ``Review Changes with AI'' which invoked the AI chat where they received feedback on best practices and code quality in terms of criteria, including modularity, efficiency, and robustness.

\end{itemize}

Due to proprietary constraints, complete system prompts could not be disclosed. However, relevant prompt segments are provided in supplementary materials~\cite{supplementary2025}.

\subsection{Contextual Inquiries}\label{sec:contextual}

We conducted contextual inquiries~\cite{holtzblatt1997contextual} with seven professional developers at JetBrains to refine approach described above. 

\subsubsection{Procedure.} Participants were chosen to be professional developers (more than 1 year of experience), regular (at least once per week) users of the IDE under study, who frequently (daily) used in-IDE AI features. Each remote session lasted approximately one hour and was both recorded and transcribed. 
Participants interacted with the early prototype described in Section \ref{sec:initial-prototype}, addressing code quality issues in any familiar project of their choosing. The setting characterized their daily work, with participants elaborating on their thought processes aloud.
We concluded each session with a semi-structured interview to gather feedback about participants' impressions, preferences, and suggestions for improvement. The script and transcripts of contextual inquiries are included in the supplementary materials~\cite{supplementary2025}. Recordings were transcribed, and through collaborative discussion, the first two authors extracted prominent and common themes to inform the refinement of the early prototype.

\subsubsection{Analysis of Themes.} Combining these insights with existing literature, we identify four Design Principles (\textbf{DPs}) that guided the refinement for our resulting implementation of ProAIDE. 

\textbf{DP1. Timely, proactive AI assistance} emphasizes anticipating developer needs without causing disruptions. In prior work, periodic suggestions had been shown to interrupt workflow and reduce focus~\cite{chen2025needhelpdesigningproactive, zhao2025codinggenieproactivellmpoweredprogramming}. To address this, our initial prototype triggered suggestions at two code iteration stages (Figure~\ref{fig:proactive_timing}): formulating needs and executing their ideas.
However, participants of the contextual inquiries recommended including triggers at task boundaries later in their workflow, such as after commits. One participant noted: \textit{``I separate writing code and optimizing it. When I optimize, I want to do it later. Reviewing should happen after I've written something and feel it's ready for review.''} [Pilot2]. We returned to this distinction between task stages and boundary-triggered interventions in our analysis of how developers engaged with proactive suggestions across workflow phases (see Section~\ref{sec:results}).

\textbf{DP2. Contextually relevant suggestions} are important for ensuring that the assistant supports, rather than interrupts, existing in-IDE workflows. 
Instead of general-purpose proactive suggestions~\cite{chen2025needhelpdesigningproactive, zhao2025codinggenieproactivellmpoweredprogramming}, we specifically targeted code-quality improvements. 
These improvements spanned optimizations based on best practices and general code quality criteria, tailored to the developer's context. Whereas the LLM underlying the AI assistant by default had access to tools to dynamically fetch code artifacts across the repository, we augmented its context with potential problems detected by the IDE (e.g. language servers) and the relevant portions of code the user was actively working on. In case of interventions on commits, the corresponding git diffs were provided.
Participants of the contextual inquiries appreciated that ProAIDE inferred their intent without requiring explicit prompts or manual context augmentation. However, they occasionally perceived suggestions as redundant, overly generic or not aligned with their initial intent. 

\textbf{DP3. Explainability and transparency} of proactive suggestions are crucial to build developer trust, particularly in industrial settings. 
Earlier work had highlighted how over-reliance and subsequent limited understanding of proactive AI suggestions could lead to long-term maintenance and extensibility issues~\cite{pu2025assistance}.
Addressing this, our design included clear explanations of suggestions and rationales behind AI intervention. Additionally, an in-chat context panel provided transparency into the AI's working context (e.g. relevant files, commits or other manual attachments from the user).
Participants valued the ability to preview changes in the editor when receiving patches, but requested inline diffs within snippets themselves, without them being integrated directly. 
Suggestions felt more trustworthy when users could inspect edits before applying them: \textit{``Maybe smaller suggestions or steps would help. Right now it feels too risky to accept all at once.''} [Pilot3]. Hence, offering multiple small patches was preferred over single large edits.

\textbf{DP4. Preserving user control} balances proactive assistance with the developer's autonomy in decision-making. From the contextual inquiries, it became evident that automatic invocation of the AI chat upon intervention points was seen as intrusive. Developers preferred lightweight, in-editor pop-ups to confirm chat invocation, noting that suggestion rejection often stemmed from task switching rather than disapproval. We found that signaling AI activity close to the user's current interaction scope within the IDE was necessary to minimize distraction. One participant commented: \textit{``I would expect some confirmation before running a new request to the chat, otherwise it would be quite unexpected.''} [Pilot5].

\subsection{ProAIDE}\label{subsec:system}
Based on findings from contextual inquiries with the early prototype, we refined ProAIDE to address the identified issues as described above (see Figure~\ref{fig:ui} for the resulting user interface):
\begin{itemize}
    \item When intervening on \textit{ambiguous prompts} and \textit{declined AI edits}, we additionally augmented the assistant's context with the corresponding user's request, to enhance alignment with their initial intent. We also refined the system prompt to avoid suggestions that may be repetitive given previous attempts (as inferred from the user request or code context).
    \item We implemented lightweight visual cues (see Figure~\ref{fig:ui}) with optional expansion directly in the IDE near the user's current cursor position (a), providing users with the option to invoke the chat panel (b), where the AI suggestions would be displayed. Each suggestion contained a concise title (b1), a brief explanation of the issue and the rationale behind the improvement (b2), and a relevant code snippet (b3). 
    \item We added an explicit option to apply suggested patches from within the AI chat, rather than directly displaying inline diffs within the editor. This on-demand model gave developers full control over the integration of suggestions while avoiding unnecessary computational cost for LLM requests in the background to apply these patches.
    
\end{itemize}

Our iterative design process resulted in an enhanced version of ProAIDE to be used in our systematic in-the-wild study of developer interactions with proactive AI assistance across authentic workflow contexts, which we describe next.
\section{In-the-wild Study Design}\label{sec:study}
To investigate developer interaction patterns with proactive AI assistance in real-world professional contexts, we conducted an in-the-wild study that employed a mixed-methods approach, combining quantitative telemetry data with qualitative survey responses. 
The study was conducted in line with institutional ethical standards, adhering to the values and guidelines outlined in the ICC/ESOMAR International Code~\cite{iccesomar}. 
The research questions we aim to address are as follows:
\begin{itemize}
    \item \textbf{RQ1.} How do developers \emph{interact} with in-IDE proactive AI suggestions across workflow stages?
    \item \textbf{RQ2.} How do developers \emph{experience} the interaction with in-IDE proactive AI suggestions?
    \item \textbf{RQ3.} How do developers \emph{perceive} the impact of in-IDE proactive AI suggestions on their work?
\end{itemize}

\subsection{Participants}

We recruited participants through JetBrains' research panel using a comprehensive screening survey, requiring active use of the Fleet IDE and the JetBrains AI assistant~\footnote{JetBrains AI Assistant: \url{https://www.jetbrains.com/ai-assistant/}. This product is a general-purpose in-IDE AI assistant that integrates code completion, AI chat, code generation, and documentation features, capturing a diverse range of interaction patterns with AI inside IDEs.} under study, regular AI tool usage (at least weekly), and professional development experience. The screening process yielded 29 qualified participants who met all selection criteria. According to the daily and post-study surveys, 15 participants were actively engaging with the tool. The number of participants is in line with prior studies of proactive AI~\cite{pu2025assistance} and in-the-wild studies of software developers~\cite{meyer2014software,meyer2017work,latoza2010developers}. 
All participants indicated familiarity with JetBrains' products \emph{and} in-IDE AI tools, minimizing potential confounding variables. Participants were required to have regular experience with AI tools in day-to-day development work. Out of the participants, 80\% indicated using AI tools multiple times daily, with 66.7\% using GitHub Copilot and 46.7\% using JetBrains AI Assistant.
The participant pool consisted of 93.3\% identifying as Developer/ Programmer/ Software Engineer roles, and 33.3\% as DevOps/ Architects. 
Regarding professional experience, 46.7\% indicated having over 11 years of coding experience and 26.7\% having 6-10 years of experience.
In terms of primary programming languages used by the participants, Python (46.7\%), Kotlin (40\%), and TypeScript (26.7\%) were the most common.
Full demographic details are in Table~\ref{tab:participant-ai-usage}.

\begin{table}[tbp]
\centering
\caption{Participant Demographics}
\label{tab:participant-ai-usage}
\small 
\setlength{\tabcolsep}{3pt}
\begin{tabular}{@{}cllcr@{}}
\toprule
\textbf{ID} & \textbf{Role} & \textbf{Languages} & \textbf{Exp.} & \textbf{AI Use Freq.} \\
\midrule
P1  & Dev, DevOps     & C\#, Groovy, Py     & 6--10y  & Multi/day \\
P2  & Dev             & Kt, TS              & 11--16y & 1/week \\
P3  & Dev, Researcher & Dart, Py, Shell     & 3--5y   & Multi/day \\
P4  & Dev             & C\#                 & 16+y    & Multi/day \\
P5  & Dev, DevOps     & Dart, Kt, Py        & 3--5y   & Few/week \\
P6  & Dev             & Java, Kt, SQL       & 11--16y & Multi/day \\
P7  & Dev, Analyst    & Py, SQL, TS         & 6--10y  & Multi/day \\
P8  & Dev             & Go, JS              & 11--16y & Multi/day \\
P9  & Architect       & Go, Kt, Py          & 16+y    & Multi/day \\
P10 & Dev             & HTML, JS, Kt        & 1--2y   & 1/day \\
P11 & Dev             & C\#, TS             & 3--5y   & Multi/day \\
P12 & Dev             & Dart, Rust, Swift   & 6--10y  & Multi/day \\
P13 & Dev, ML Eng     & Py                  & 6--10y  & Multi/day \\
P14 & Dev, Architect  & Go, PHP, TS         & 16+y    & Multi/day \\
P15 & Dev, Architect  & Kt, Py              & 11--16y & Multi/day \\
\bottomrule
\end{tabular}
\end{table}

\subsection{Data collection procedure}

The study followed a structured timeline spanning four weeks from 7 May 2025 to 3 June 2025. 
Participants were recruited on a rolling basis and enrolled in batches. 
Eligible participants, identified through a screening survey, received an invitation to start the study in the following week. The core study period required five active days of Fleet IDE use. 
This duration balanced capturing interactions with ProAIDE beyond initial novelty~\cite{kefalidou2016encouraging,meyer2017work} with the practical constraints of professional developers. 
To further accommodate participant schedules, we allowed a ten-day window for completing these five active days. Participants received a reminder midway through the study and a final survey prompt at the end of the participation window. This design provided flexibility for participants while maintaining structured, consistent data collection. Supplementary materials~\cite{supplementary2025} provide detailed texts used for that.

\subsubsection{Capturing in-the-wild usage and daily surveys.} During the study, participants engaged in self-directed programming tasks representative of their routine professional activities (\emph{i.e.} ``worked as usual''). There were no constraints on programming languages, frameworks, or project types for this reason. With participants' informed consent, we collected anonymous IDE usage statistics as telemetry data. Importantly, we did not collect personal data, source code, file names, or full inputs and outputs exchanged between the IDE and LLMs.
To prevent fraudulent participation, in-IDE notifications prompted participants to complete short daily surveys. These surveys appeared five minutes after interactions with any proactive feature, limited to once per IDE session. Surveys captured the specific proactive feature used, the relevant development task, and perceptions of feature behavior, along with an open-ended question for additional feedback. Daily surveys also helped track participation, as telemetry data were anonymous and thus not linked to individual participants.

\subsubsection{Post-study survey.} After completing at least five active days and corresponding daily surveys, participants filled out a post-study survey. This survey included Likert-scale items from the System Usability Scale (SUS)~\cite{lewis2018system}. Additional questions evaluated participants' perceptions of the assistant's reliability, intent alignment, intrusiveness, and impact on efficiency. Two open-ended questions encouraged participants to describe specific effects on code quality and suggest improvements to better support their workflow.

\subsection{Data Analysis}

We applied both quantitative and qualitative analysis methods to address our research questions. 
To examine developer \textbf{interactions} with the proactive feature of AI assistant across workflow stages~\ref{sec:stages} (\textbf{RQ1}), we analyzed 5,732 data points and 229 interventions in telemetry logs. Specifically, we focused on three interaction types for each AI intervention: engage (user opted for viewing the suggestions in chat), dismiss (user actively closed the notification), or ignore (no user response). We processed raw telemetry events by filtering for proactive intervention triggers, then categorized subsequent user actions following each intervention. 
Descriptive statistics, including engagement patterns and temporal patterns across the five-day study period, were computed to characterize interaction behaviors for each intervention type (ambiguous prompt, declined AI edit, commit changes).
To evaluate developer \textbf{perceptions} of the assistant (\textbf{RQ2}) and its perceived \textbf{impact} (\textbf{RQ3}), we mainly analyzed post-study survey responses. This included quantitative data summarized through frequency distributions for Likert-scale items, and qualitative insights from open-ended questions. Moreover, we employed a Wilcoxon signed-rank test~\cite{scipyWilcoxon} to compare interpretation times (from the moment a suggested snippet appeared in the chat interface until the user's first response to it) between proactive and reactive chat sessions. Given the moderate sample sizes and the observed distributional skewness inconsistent with normality assumptions in both conditions, the Wilcoxon signed-rank test provided robust inference without requiring the parametric assumptions necessary for t-tests. We combine the telemetry and survey data to derive contextualized insights into the usability and perceived usefulness of ProAIDE.\footnote{Survey data supporting the findings of this study may be shared upon reasonable request. Due to privacy and confidentiality concerns, telemetry data collected from in-IDE usage cannot be made publicly available.}

\section{Results}\label{sec:results}

Our analysis combines in-IDE telemetry (N=18 machines, with active usage declining to 8 machines by day five) and a post-study questionnaire with open-ended feedback (N=15).

\subsection{RQ1: Interaction Patterns Across Workflow}

\begin{figure*}[tbp]
    \centering
    \includegraphics[width=\textwidth]{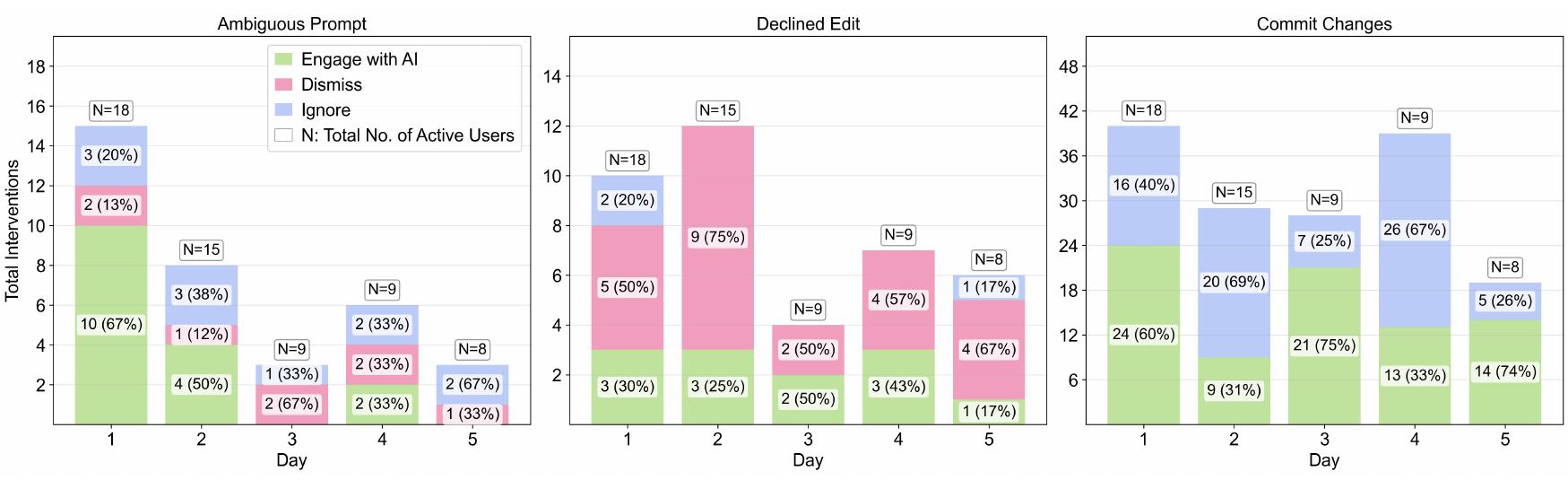}
    \caption{\textbf{Rates of interaction with proactive AI interventions across various development stages:} formulating needs (Ambiguous Prompt), executing the idea (Declined Edit) and finishing writing code (Commit Changes). Users had the option to Engage, Dismiss, or Ignore AI interventions. Active participation decreased from 18 to 8 users between Day 1 and Day 5, yet interaction patterns remained stable across intervention types. Commit Changes interventions achieve the highest engagement (52\%), while Declined Edit interventions show consistently low engagement (31\%). In total, we recorded 229 proactive AI interventions over the 5-day study period.}
    \Description{Three stacked bar charts showing interaction rates over 5 days for three intervention types. Left chart (Ambiguous Prompt, N=35 total): shows Engage (green), Dismiss (pink), and Ignore (purple) rates varying by day, with engagement around 46 percent overall. Middle chart (Declined Edit, N=39 total): shows consistently low engagement around 31 percent with high dismissal rates around 62 percent across all days. Right chart (Commit Changes, N=155 total): shows highest engagement rates around 52 percent, with active users declining from 18 on Day 1 to 9 on Day 5. Each bar shows the total number of active users (N) for that day.}

    \label{fig:editor_interventions}
\end{figure*}

To answer the \textbf{RQ1}, we analyzed 229 interactions (Figure~\ref{fig:editor_interventions}) across three workflow-triggered intervention types: (1) \textit{Ambiguous Prompt}, (2) \textit{Declined AI Edit}, and (3) \textit{Commit Changes} over the study period of five days.
Each intervention provided developers with options to either engage with the AI for code improvement suggestions, dismiss the notification, or ignore it entirely.

\begin{itemize}
    \item \textit{Ambiguous Prompt} interventions (N=35) saw moderate engagement (46\%), with 23\% dismissals and 31\% ignores. This balanced distribution may reflect a mix of situations: in some cases, developers appeared to be in an exploratory mode, open to proactive suggestions. In others, they seemed to have a clear intent in mind but preferred efficiency over detail, being surprised that \textit{``simplify was insufficient context for the AI to act upon''} [P1]. 
    
    \item \textit{Declined AI Edit} interventions (N=39) achieved the lowest engagement rates at 31\% and highest dismissal rates at 62\%. Despite signaling unmet needs, these interventions appear to feel \textit{"like advertisement"} [P3], indicating unnecessary intrusion when developers have already formulated specific task intentions. Moreover, context-switching between different AI interaction modalities appears to be causing cognitive overload, with participants expressing confusion about \textit{"when to use inline AI and chat"} [P15]. 

    \item \textit{Commit Changes} interventions (N=155) achieved highest engagement rates (52\%). This timing is based on natural workflow boundary points when developers complete logical units of work and are \textit{"still in the mindset of wanting to submit good code"} [P5]. This suggests that developers, having finished writing code and entered an evaluative mindset, experience reduced cognitive switching costs and thus are more receptive to proactive AI. 
\end{itemize}

An analysis of interaction patterns over the study period revealed stability across all intervention types (See Fig.~\ref{fig:editor_interventions}). 
For \textit{Commit Changes} interventions, the consistent engagement rates throughout the study period indicate that the utility of boundary-point interventions is not diminished by repeated exposure. 
The daily survey feedback confirmed that having AI review changes after commits remains helpful, with users expressing a desire to \textit{``continue using it once the feature becomes available''} [P5]. 
In contrast, engagement rates for \textit{Declined AI Edit} interventions remained consistently low, suggesting that the poor reception was not necessarily due to an initial learning curve, but rather reflects a more fundamental disruption to their workflow.
Moreover, 8 out of 18 participants chose to continue using ProAIDE beyond the mandatory 5-day study period, suggesting that its value extended beyond initial curiosity. This sustained engagement indicates that the proactive feature was not merely perceived as a novel experimental tool, but had begun to integrate meaningfully into participants' regular workflows.

\rqsummary{RQ1}{Proactive AI's utility depends on developers' cognitive context and thus readiness to receive assistance. The most accepted interventions occur at natural workflow boundaries when developers are in the mindset of reviewing submitted code, while interventions during focused implementation tasks often feel disruptive and intrusive.}

\subsection{RQ2: User Experience} 

\begin{figure*}[tbp]
    \centering
    \includegraphics[width=\textwidth]{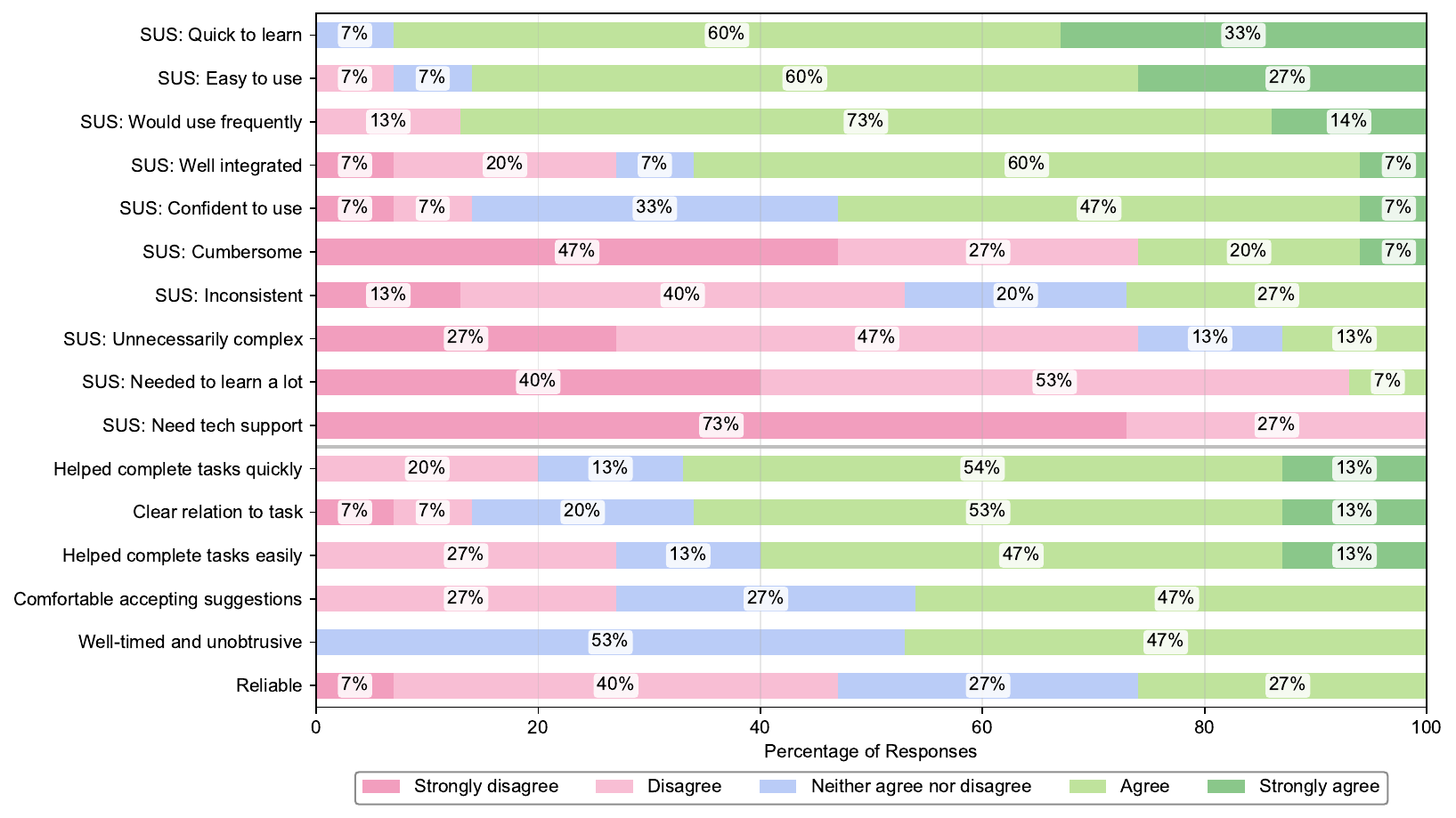}
    \caption{\textbf{System Usability Scale and user experience responses} from 15 participants who completed the entire study period. Results show above-average usability ($SUS = 72.8$).}
    \Description{Horizontal stacked bar chart showing survey responses from 15 participants across 16 items. Items are grouped into SUS questions and custom experience questions. Each bar shows percentage distribution across five categories: Strongly disagree (dark red), Disagree (light red), Neither agree nor disagree (gray), Agree (light green), and Strongly agree (dark green). Key findings: 93 percent found the system quick to learn, 87 percent found it easy to use, 66 percent said it helped complete tasks quickly, but only 27 percent rated it as reliable. The overall SUS score is 72.8 out of 100.}
    \label{fig:survey-combined}
\end{figure*}

Having identified when proactive AI interventions are most interacted with, we now examine how developers experienced these interventions (\textbf{RQ2}). The findings below are primarily based on analysis of qualitative post-study survey responses (Figure~\ref{fig:survey-combined}).

\subsubsection{ProAIDE was Easy to Use and Learn}

Participants generally rated ProAIDE positively in terms of usability, reflected by an above-average SUS score of 72.8 out of 100 (95\% CI: [64.1, 81.5]). This indicates good usability, though with room for improvement, suggesting users hold favorable but not strongly promotional views~\cite{sauro2016quantifying, clark2021empirical}. The majority found the system quick to learn (93\%), likely due to its seamless integration within the already familiar IDE environment. Additionally, participants considered the proactive feature easy to use (87\%). Explicit confirmation mechanisms were particularly appreciated, with one participant noting the system was \textit{"even better than some other IDEs [...] offering custom UI in form of buttons when asking for confirmation"} [P12].
Furthermore, the transparency in AI presence contributed to helping developers feel like they \textit{“always knew what the suggestions were talking about”} [P3].

\subsubsection{Developers Had Varying Needs and Preferences}

In terms of how proactive interventions were triggered, many expressed a desire for control, as [P5] noted: \textit{"I prefer to use hotkeys where possible or a command that can be used"}. Conversely, some preferred greater AI autonomy, with another participant stating \textit{"[They] would like to immediately move context to 'Chat' without even asking me, since [they] didn't accomplish what [they] tried to do"} [P15].
Moreover, opinions regarding the degree of proactiveness and granularity of suggestions differed. Some participants indicated that \textit{"it would be good if the 'Review Commit' option could be triggered automatically on any commit"} [P5], or anchored to \textit{"any changes made in git"} [P10]. 
Others found intervention frequency excessive, stating that suggestions were \textit{"not always necessary [...] dismissing the dialog is annoying"} [P15]. Regarding suggestion granularity, some participants sought detailed, \textit{"fine-grained comments in code and documentation"} [P3], while others preferred concise suggestions, stating \textit{"no need for explanation text"} [P15]. 
Also, participants also wished to select suggestion types explicitly, such as those related to \textit{"performance, readability, security"} [P4].

\subsubsection{Alignment Between AI and Mental Model Was Key}

An important factor influencing developers' receptiveness to proactive suggestions was how well the AI's behavior aligned with their expectations and mental models. Survey responses indicated that 66\% of participants easily understood how the suggestions related to their intended tasks.
One participant noted, \textit{``In most cases, suggestions were exactly what I wanted''} [P6]. 
Interestingly, one participant emphasized the role of developer context, stating, \textit{``I did not feel that any suggestion was misleading [...] so perhaps the coder's perception and friction with the code they are writing is an important factor here.''} 
This suggests that perceived alignment may depend not only on the AI but also on the developer's experience with the task at hand.

\subsubsection{Perceived Utility was Shaped by the AI's Contextual Understanding}

The usefulness of proactive AI suggestions was strongly influenced by the AI's ability to understand the developer's code context.
However, only 27\% of participants rated AI chat suggestions as reliable, and just 47\% felt comfortable accepting them.
In fact, the perceived reliability diminished when intent behind design choices was not understood; in some cases \textit{``repetitiveness was necessary and unavoidable for the specific functionality intended to achieve''} [P3]. Others experienced frustration when the \textit{``AI failed to account for domain-specific syntax and patterns''} [P13].
We note that LLM training data here plays a substantial role, as P3 observed: \textit{``at least half of the times this happened [...] other LLM chats [...] would get it wrong in their first try as well.''}
Another described the effort required to manually adjust inaccurate suggestions: \textit{``The suggestions were sort of helpful, but messed up the code file. I manually copied suggestions, undid changes, then pasted suggestions into the correct place''} [P4], disrupting their workflow.

\subsubsection{Specific Friction Points}

Beyond contextual understanding, session length also affected reliability. Participant [P7] observed that ``if a particular chat session gets long, then the model would start glitching out'', entering loops where it would ``list out the entire plan as is and not actually implement the plan''. 
This degradation over extended interactions forced developers to start fresh sessions, disrupting workflow continuity. 
Finally, some participants noted that commit review suggestions sometimes restated changes rather than critiquing them. Participant [P13] provided feedback that ``instead of reviewing my commit, it summarizes the changes I made and presents them as its `suggestions' ''.

\rqsummary{RQ2}{Developers rated the proactive feature as easy to use (SUS = 72.8), expressing ease of learning and good integration. 
However, preferences varied widely regarding user control, degree of proactivity, and granularity of suggestions.
Perceived usefulness was influenced by how well the AI's suggestions aligned with developers' mental models and workspace context.}

\begin{figure}[tpb]
    \centering
    \includegraphics[width=\linewidth]{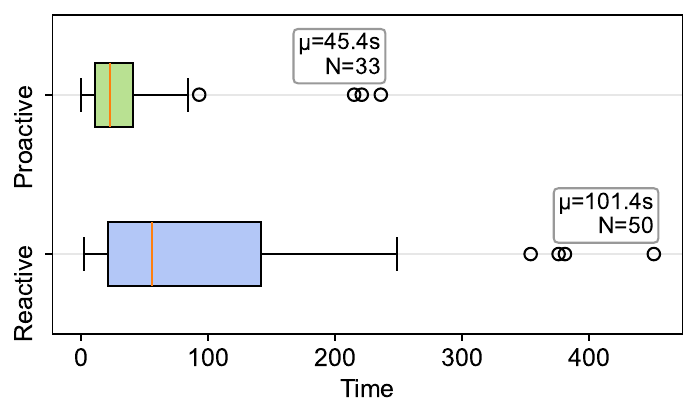}
    \caption{Interpretation time comparison between proactively triggered chat sessions ($\mu=45.4s$, $N=33$) and reactively triggered ones ($\mu=101.4s$, $N=50$). Proactive suggestions demonstrate significantly faster interpretation times ($p=0.0016$, Wilcoxon signed-rank test), suggesting improved efficiency when AI assistance is tailored to the user's workflow context.}
    \Description{Box plot comparing interpretation times between proactive and reactive chat sessions. The proactive condition (top) shows a median around 45 seconds with mean of 45.4 seconds and N=33 sessions, with a compact distribution. The reactive condition (bottom) shows a median around 100 seconds with mean of 101.4 seconds and N=50 sessions, with a wider distribution extending to approximately 400 seconds. The difference is statistically significant with p=0.0016 using Wilcoxon signed-rank test, indicating proactive suggestions are interpreted significantly faster.}

    \label{fig:rq3}
\end{figure}

\subsection{RQ3: Impact of Proactive Suggestions}

While RQ2 focused on the experience of using ProAIDE, we now turn to RQ3, which investigates how its suggestions influenced developers' perception of impact on their work. The study reveals that proactive AI assistance generates overall positive effects on perceived productivity.
We find that the majority of participants indicated that using the proactive suggestions helped them complete tasks more quickly (66\%) and easily (60\%), as seen in Figure~\ref{fig:survey-combined}. Interestingly, from qualitative responses, we find that developers engaged with ProAIDE for various underlying purposes.

\subsubsection{The Proactive AI Served Exploratory Purposes}

Our ProAIDE served as a code quality gate, with participants appreciating its role as a second pair of eyes for identifying overlooked issues or general enhancements. P3 noted: \textit{``There were a few suggestions around safety ... an aspect that actually is really important in order to catch senseless errors when speed coding and did improve my code significantly.''} 
Similarly, P5 mentioned how \textit{``[They] didn't catch that loading the asset might throw an error as [they were] following an implementation guide from the library docs.''}
Participants also seemed to get an intuition for the AI's strengths over time: \textit{``the AI is helpful for identifying repetitive code and extracting it to improve modularity. It also handles typing well.''} [P13]. 

\subsubsection{The Proactive AI Served Accelerative Purposes}

Beyond exploratory purposes, we observed that ProAIDE served accelerative purposes by alleviating the manual burden of context management and switching between different AI modalities, especially on \textit{Declined AI Edit} interventions.
While developers generally rejected proactive suggestions after declining AI edits, some participants appreciated that the system automatically escalated to the AI chat with relevant context (e.g. when inline AI failed to produce the desired result), \textit{``offering proactive code editing while [they were] writing code so that [they wouldn't] have to open a chat window or go somewhere else to fill in what's missing''} [P12].

Interestingly, users spent significantly less time interpreting suggestions in proactively-triggered chat sessions ($\mu=45.4s$, $N=33$) compared to reactively-triggered sessions ($\mu=101.4s$, $N=50$), as shown in Figure~\ref{fig:rq3}.

We compared interpretation times (\emph{i.e.,} the time between the moment a suggested patch or snippet was presented in the chat until the user responded using either ``Accept'' or ``Copy'') between proactive and reactive chat sessions using data from the same set of $18$ participants. Given the paired nature of the data and the observed distributional skewness (see Figure~\ref{fig:rq3}), we employed the Wilcoxon signed-rank test as a non-parametric alternative to the paired $t$-test. To address the unequal group sizes (50 reactive vs. 33 proactive sessions), we created balanced subsets by randomly sampling 33 observations from each condition. The test revealed a statistically significant difference ($W = 109.00$, $p = 0.0016$) with a medium-to-large effect size ($r = 0.533$).
These findings suggest that proactively delivered suggestions, whether accepted as patches or adapted as snippets, were substantially easier and faster for developers to interpret than reactively requested ones.

Examining the types of suggestions reveals differences in user engagement. In the reactive condition, users encountered relatively more exemplary code snippets (21/50; 42\%) than in the proactive condition (5/33; 15\%), where patches dominated. This discrepancy suggests that reactive use was more information-seeking, while proactive suggestions provided more actionable code. 
This distinction likely contributed to the reduced interpretation time observed for proactive suggestions, as users were presented with ready-to-apply patches that generated diffs directly in the editor. In contrast, exemplary snippets were displayed only as in-chat code fragments, likely requiring more effort to interpret and adapt.

\rqsummary{RQ3}{Developers engage with proactive AI for both exploratory and accelerative purposes, achieving high perceived productivity and workflow efficiency. In fact, we observe significantly lower interpretation times for proactive suggestions compared to reactive AI use.}

\section{Discussion}

\subsection{Design Implications for in-IDE Proactive AI}

Our study provides empirical evidence for how developers interact with proactive AI assistance in authentic professional workflows, revealing timing-dependent patterns that inform the design of context-aware coding assistants. While prior work explored proactivity in controlled environments and limited contexts, our in-the-wild study reveals how real-world developers engage with and evaluate proactive interventions over time, spanning various programming languages and frameworks. We also revisit findings from earlier studies, contrasting them with our observations based on real-world behavior. Building on these insights, we offer design implications and recommendations grounded in the four design principles introduced in Section~\ref{sec:design-principles}.

\textbf{DP1. Timely, proactive AI assistance.}
Our findings reveal that timing is a crucial aspect impacting developer engagement with proactive AI. Unlike static recommendation systems, LLM-based proactive assistants must dynamically align their understanding with developer mental models through contextual reasoning. Workflow boundary interventions (e.g., after commits) were consistently well-received, due to natural blending into developer workflows. This reinforces prior findings from Pu et al.~\cite{pu2025assistance}, now ecologically validated in a production environment.
On the other hand, mid-task interventions (e.g., after declined edits) were frequently dismissed or ignored.
This pattern aligns with Iqbal and Bailey's Index of Opportunity framework~\cite{iqbal2005towards}, which demonstrates that mental workload decreases at task boundaries, creating natural windows for interruption. Post-commit moments represent such boundaries: developers have externalized their working memory into the commit, freeing attentional resources for evaluating new information. Mid-task interventions, by contrast, compete for cognitive resources engaged in active implementation. This theoretical lens suggests that receptivity to proactive AI depends on the underlying cognitive state that different workflow moments reflect.

\textit{Implication and Recommendation:} Developers' workflow context and cognitive state must guide when interventions are triggered to ensure effectiveness.
As such, anchor proactive suggestions to clear task boundaries (e.g., failed builds or test runs). For mid-task interventions, we recommend tuning down the AI's prominence in the IDE or providing configuration options to respect individual workflows.~\footnote{Examples of such customization mechanisms include custom slash commands and hooks in the CLI-based AI coding tool Claude Code~\cite{anthropic2024claudecode}.} Overall, we highlight the need for more accurate intent modeling to adapt triggers dynamically based on developer behavior and context. Reinforcement learning (RL), for instance, can be used to learn optimal interruption moments to balance relevance and flow disruption~\cite{Ho2020}. Our study contributes the empirical foundation such approaches require: the engagement and dismissal outcomes we observed provide ground truth labels for developer receptivity, which could serve as reward signals in future RL formulations. On the other hand, in-IDE telemetry events offer candidate state representations. However, learning such policies requires larger-scale deployments with appropriate privacy safeguards, which we leave to future work.

\textbf{DP2. Contextually relevant suggestions.}
Developers valued suggestions that reflected their current task, responding positively when suggestions felt well-aligned.
Moreover, our findings support \citet{chen2025needhelpdesigningproactive}'s framing of proactive AI as a ``second pair of eyes'' to catch overlooked issues.
Nonetheless, we identify that the perceived value of suggestions diminished when they were perceived as irrelevant, for instance, due to unawareness of domain-specific patterns, which can be attributed to both context integration and LLM training data. 

\textit{Implication and Recommendation:} Suggestion quality is largely influenced by the AI's understanding of developer intent and their workspace, highlighting the importance of enhanced integration with workspace context.
As such, adopt more effective heuristics for integrating in-IDE context sources, prioritizing signals that reflect the developer's immediate focus i.e. \textit{``possibly using nearby open tabs or recently edited files''} [P3], ideally grounding these signals in a broader semantic representation of the codebase through repository indexing. 
Notably, recent standardized protocols for connecting LLMs to external data sources such as Anthropic’s MCP~\cite{anthropic2024context} exemplify a shift toward structured, tool-mediated context retrieval enabling real-time personalized augmentation of model context.

\textbf{DP3. Explainability and transparency.}
Transparency in the AI's in-IDE presence was essential for establishing reliability. Developers needed to understand not just what the AI suggested, but why it triggered at that moment and how it interpreted the code context. This differs from traditional recommendation systems, where the underlying logic is more predictable.
Developers appreciated concise rationales, previewable changes, and minimal in-editor cues that respected their workflow.
Notably, while Pu et al.~\cite{pu2025assistance} found that flexible interaction scopes improved transparency, our findings show they can cause distraction due to context switching. This highlights the need for careful coordination of concurrent AI features in real-world IDEs to reduce cognitive load.
To further enhance trust, participants indicated that it \textit{``would be helpful if each suggestion came with a confidence level''}.

\textit{Implication and Recommendation:} Proactive suggestions must be understandable at a glance, predictable in behavior, and verifiable before application.
As such, enhance developer decision-making by increasing transparency, e.g., showing clear inline diffs when presenting patches (i.e. before application onto the editor) or confidence indicators with suggestions.

\textbf{DP4. Preserving user control.}
Control over when and how to engage with AI suggestions was a recurring theme. In fact, preferences varied across individuals; some preferred more autonomy, while others favored greater manual control via
hotkeys or menu actions. Beyond interaction modalities, users expressed a desire to have control over the type and granularity of suggestions.

\textit{Implication and Recommendation:} To accommodate diverse user workflows and preferences, adaptive human-AI interaction strategies are needed.
As such, combine user-configurable settings for intervention timing, frequency, and suggestion types with adaptive assistance that learns from both explicit feedback and implicit behavioral signals to provide personalized support.

\subsection{Threats to Validity}

While our findings provide strong real-world insights into proactive AI use, several limitations must be acknowledged.

\textbf{External Validity.} Participants were recruited from JetBrains' research panel and were familiar with its IDEs and AI tools, which allowed us to study ecologically valid use cases. However, the findings may not generalize to developers unfamiliar with these tools or those with limited AI experience. While the findings likely generalize to similar in-IDE AI assistance tools, future studies should broaden the participant pool to assess representativeness across diverse contexts.
Due to the limited sample size, we did not stratify participants by role, experience level, or language. Some findings in our study may be shaped by language-specific workflows or by domain-specific factors. For languages with limited tool support or different compilation models (e.g., C++), tailored designs or fallback strategies may be necessary to achieve similar outcomes. Future studies should examine whether timing-dependent receptivity patterns vary systematically across developer categories.

\textbf{Construct Validity.} Participants interacted with a full-featured AI assistant, of which our proactive feature was one component. As such, post-study perceptions may have been shaped by the assistant as a whole. We see this as a strength of ecological integration, yet it introduces ambiguity in interpreting usability scores. Still, qualitative data confirms that the proactive feature contributed to the observed user experience, making these responses relevant.
Controlled A/B testing may complement in-the-wild studies by isolating specific feature effects.
Our assessment of proactive suggestions' impact relies on self-report rather than objective metrics. Preserving participant privacy in enterprise settings precluded collecting source code or LLM outputs, preventing direct measurement of code quality improvements. Future work could address this through controlled studies with open-source repositories where code artifacts and peer review outcomes can be analyzed, or by integrating privacy-preserving quality signals into telemetry with participant consent.

\textbf{Internal Validity.} Participants were exposed to all intervention types, but engagement varied by individual workflow, IDE usage intensity, and day-to-day conditions. As telemetry data were collected anonymized and separately from surveys, we could not link behavioral data to subjective perceptions at the individual level. While both datasets are independently reliable, they cannot be precisely mapped to the same individuals. Moreover, telemetry includes data from all participants who engaged with the system, even if they did not complete the full five-day protocol (N=18 machines with active usage declining to 8 machines by day five). On the contrary, survey analyses were restricted to participants who met the five-day filled daily survey protocol criteria (N=15). This preserves privacy and maximizes available data but may have introduced noise.

\textbf{Statistical Conclusion Validity.} The sample size, variability in daily engagement, and non-uniform exposure to intervention types limit the power of inferential statistical tests. Where applicable, we use non-parametric tests to account for data specifics. While our findings point to meaningful patterns, results should be interpreted with caution due to the limited power.

\section{Conclusion}

We investigated how professional developers interact with proactive AI assistance in authentic workflow contexts through a five-day in-the-wild study with 15 developers within a production IDE. 
We demonstrated that \textit{timing} is key: suggestions presented at natural workflow boundaries (e.g., immediately post-commit) garnered the highest engagement rates, whereas mid-task interventions were often dismissed.
Importantly, we observed sustained engagement patterns over the course of the study, indicating a potential for long-term integration of proactive assistance into everyday development workflows.
Participants consistently valued ProAIDE's ability to surface overlooked issues, particularly when it functioned as a lightweight code quality gate.
These interventions also supported perceived productivity and efficiency, given the observation that proactive suggestions were interpreted and applied significantly faster than reactive ones.
Our findings bridge the gap between conceptual AI prototypes and production-ready IDE integration. We offer design insights and empirical evidence for embedding proactive AI into daily software engineering workflows, emphasizing adaptive support and user-configurable behavior to lay the groundwork for future research.
While our five-day study revealed adoption patterns, longer-term research is needed to understand retention, evolving perceptions, and lasting impact.

\section{GenAI Usage Disclosure}
Portions of this manuscript were revised using OpenAI’s ChatGPT for grammar and clarity improvements. The authors reviewed and edited all suggestions and take full responsibility for the content.

\begin{acks}
This work was supported by JetBrains as part of the AI for Software Engineering (AI4SE) collaboration with Delft University of Technology.
\end{acks}

\bibliographystyle{ACM-Reference-Format}
\bibliography{main}

\end{document}